\documentclass{aa}
\usepackage{graphicx}
\usepackage{txfonts}
\usepackage[switch]{lineno}
\usepackage{txfonts}
\usepackage{natbib}
\usepackage{graphicx, xspace}
\usepackage{multirow}
\usepackage{ amssymb }

\newcommand{\vsini}{$v \sin i$\xspace}
\newcommand{\nm}{\,nm\xspace}

\newcommand{\num} {$\nu_{\rm max}$\xspace}
\newcommand{\logg} {$\log g$\xspace}

\newcommand{\Kepler} {\textit{Kepler}\xspace}

\newcommand{\Hermes} {\textsc{Hermes}\xspace}
\newcommand{\hermes} {\textsc{Hermes}\xspace}
\newcommand{\Mercator} {\textsc{Mercator}\xspace}

\newcommand{\KIC}[1]{{KIC\,#1\xspace}}

\newcommand{\dex}{\,dex\xspace}

\newcommand{\gssp}{{\sc gssp}\xspace}

\def\teff{$T_{\mathrm{eff}}$\xspace}

\def\dn1{$\delta\nu_{01}$\xspace}
\def\dn2{$\delta\nu_{02}$\xspace}
\def\sun{\hbox{$_\odot$}\xspace}

\newcommand{\prot}{P$_{\rm rot}$\xspace}
\newcommand{\Prot}{P$_{\rm rot}$\xspace}

\usepackage[dvips]{color}
\newcommand{\blu}{\textcolor{blue} }

\newcommand{\new}[1]{\textbf{\pink{#1}}}
\newcommand{\newer}[1]{\blu{#1}}

\newcommand{\referee}[1]{\textbf{\blu{#1}}}

\newcommand{\Figure}[1]{Figure\,\ref{#1}\xspace}
\newcommand{\Table}[1]{Table\,\ref{#1}\xspace}
\newcommand{\Equation}[1]{Equation\,(\ref{#1})\xspace}
\newcommand{\Section}[1]{Section\,\ref{#1}\xspace}

\usepackage[colorlinks=True,
			linkcolor=blue,
			citecolor=blue, 
			filecolor=blue, 
			urlcolor=blue]{hyperref}

\renewcommand{\new}[1]{{#1}}
\renewcommand{\newer}[1]{{#1}}

\renewcommand{\blu}[1]{{#1}}
\renewcommand{\referee}[1]{{#1}}

\begin{document}

\title{Lithium abundance and rotation of seismic solar analogues}
\subtitle{Solar and stellar connection from  \Kepler and \Hermes observations\thanks{Based on observations made with the NASA \Kepler space telescope and the \Hermes spectrograph mounted on the 1.2\,m \Mercator Telescope at the Spanish Observatorio del Roque de los Muchachos of the Instituto de Astrof\'isica de Canarias.}}

\titlerunning{Lithium Abundance and Rotation of seismic Solar Analogues.}
\authorrunning{Beck et al.}


\author{P.\,G.~Beck\inst{1} 
\and J.-D.~do\,Nascimento Jr.\inst{2,3}
\and T.~Duarte\inst{2} 
\and D.\,Salabert\inst{1}
\and A.~Tkachenko\inst{4} 
\and S.~Mathis\inst{1},\\
 S.~Mathur\inst{5} 
\and R.\,A.~Garc\'\i a\inst{1}
\and M.~Castro\inst{2}
\and P.\,L.~Pall\'e\inst{6,7}
\and R.~Egeland \inst{8,9}
\and D.~Montes\inst{7}
\and O.~Creevey\inst{11}, \\
 M.~F. Andersen\inst{12}
\and D.~Kamath\inst{4}
\and H.~van\,Winckel\inst{4}  
} 
\mail{paul.beck@cea.fr}

\institute{Laboratoire AIM, CEA/DRF - CNRS - Univ. Paris Diderot - IRFU/SAp, Centre de Saclay, 91191 Gif-sur-Yvette Cedex, France 
\email{paul.beck@cea.fr}
\and Departamento de F\'isica, Universidade Federal do Rio Grande do Norte, 59072-970 Natal, RN, Brazil 
\and Harvard-Smithsonian Center for Astrophysics, 60 Garden Street, Cambridge, MA 02138, USA
\and Instituut voor Sterrenkunde, KU Leuven, B-3001 Leuven, Belgium 
\and Space Science Institute, 4750 Walnut street Suite 205, Boulder, CO 80301, USA 
\and Instituto de Astrof\'{\i}sica de Canarias, E-38200 La Laguna, Tenerife, Spain 
\and Departamento de Astrof\'{\i}sica, Universidad de La Laguna, E-38206 La Laguna, Tenerife, Spain 
\and High Altitude Observatory, National Center for Atmospheric Research, P.O. Box 3000, Boulder, CO 80307-3000, USA 
\and Department of Physics, Montana State University, Bozeman, MT 59717-3840, USA 
\and Dpto. Astrof\'{\i}sica, Facultad de CC. F\'{\i}sicas, Universidad Complutense de Madrid, E-28040 Madrid, Spain 
\and Laboratoire Lagrange, Universit\'e de Nice Sophia-Antipolis, UMR 7293, CNRS, Observatoire de la C{\^ o}te d'Azur, Nice, France
\and Stellar Astrophysics Centre, Aarhus University, Ny Munkegade 120, 8000 Aarhus C, Denmark 
}

\date{Received: 30 September 2016 / Accepted: 2 February 2017}
 
  \abstract
{Lithium abundance A(Li) and surface rotation are good diagnostic tools to probe the internal mixing and  angular momentum transfer in stars.} 
{We explore the relation between surface rotation, A(Li) and age in a sample of seismic solar-analogue stars and we study their possible binary nature.} 
{We select a sample of 18 solar-analogue stars observed by the NASA \Kepler satellite for an in-depth analysis. Their seismic properties and surface rotation rates are well constrained from previous studies. About 53 hours of high-resolution spectroscopy were obtained  to derive fundamental parameters from spectroscopy and A(Li). 
These values were combined and confronted with seismic masses, radii and ages, as well as  surface rotation periods measured from \Kepler photometry.} 
{Based on radial velocities, we identify and confirm a total of 6 binary star systems.  
For each star, a signal-to-noise ratio of 80$\lesssim$S/N$\lesssim$210 was typically achieved in the final spectrum around the lithium line.  
We report fundamental parameters and A(Li). By using the surface rotation period derived from \Kepler photometry, a well-defined relation between A(Li) and rotation was obtained. The seismic radius translates the surface rotation period into surface velocity.
With models constrained by the characterisation of the individual mode frequencies for single stars,  we identify a sequence of three solar analogues with similar mass ($\sim$1.1\,M\sun) and stellar ages ranging between 1 to 9\,Gyr. Within the realistic estimate of $\sim$7\% for the mass uncertainty, we find a good agreement between the measured A(Li) and the  predicted A(Li) evolution from a grid of models calculated with the Toulouse-Geneva stellar evolution code, which includes rotational internal mixing, calibrated to reproduce solar chemical properties. 
We found a scatter in ages inferred from the global seismic parameters, too large when compared with A(Li). } 
{We present Li-abundance for a consistent spectroscopic survey of solar-analogue stars with a mass of 1.00$\pm$0.15\,M\sun, and characterised through asteroseismology and surface rotation rates based on \Kepler observations. The correlation between A(Li) and \prot supports the gyrochronological concept for stars younger than the Sun and becomes clearer, if the confirmed binaries are excluded. The consensus between measured A(Li) for solar analogues with model grids, calibrated onto the Sun's chemical properties suggests that these targets share the same internal physics. In this light, the solar Li and rotation rate appear to be normal for a star like the Sun.
}


\maketitle




\section{Introduction}

\begin{table*}

\caption{Parameters found in the literature of the stars used in this study.}
\vspace{-2mm}
\centering
\tabcolsep=15pt
\begin{tabular}{lcrrrrc}
\hline\hline
\multicolumn{1}{c}{KIC} & 
\multicolumn{1}{c}{\num} & 
\multicolumn{1}{c}{M} & 
\multicolumn{1}{c}{R} & 
\multicolumn{1}{c}{P$_{\rm rot}$} &
\multicolumn{1}{c}{Age} & 
\multicolumn{1}{c}{Ref.} \\ &
\multicolumn{1}{c}{[$\mu$Hz]} & 
\multicolumn{1}{c}{[M$_\odot$]} & 
\multicolumn{1}{c}{[R$_\odot$]}& 
\multicolumn{1}{c}{[days]} &
\multicolumn{1}{c}{[Gyr]}& 
\multicolumn{1}{c}{}
\\

\hline



3241581$^\star$	& 2969$\pm$17& 1.04$\pm$0.02 & 1.08$\pm$0.10  &   26.3$\pm$2.0  & 3.8$\pm$0.6 &  1	\\

3656476$^\star$	& 1947$\pm$78&  1.10$\pm$0.03 & 1.32$\pm$0.01    & 31.7$\pm$3.5 &  8.9$\pm$0.4 &  2 	\\

4914923$^\star$	& 1844$\pm$73&  1.04$\pm$0.03 & 1.34$\pm$0.02   & 20.5$\pm$2.8 &  7.0$\pm$0.5 & 2	\\

5084157	& 1788$\pm$14&  1.06$\pm$0.13 &  1.36$\pm$0.08  &   22.2$\pm$2.8 &  7.8$\pm$3.4 & 3	\\

5774694	& 3671$\pm$20& 1.06$\pm$0.05 & 1.00$\pm$0.03   &  12.1$\pm$1.0 &  1.9$\pm$1.8 & 3 \\

6116048$^\star$	& 2098$\pm$84&  1.05$\pm$0.03 & 1.23$\pm$0.01   & 17.3$\pm$2.0 &  6.1$\pm$0.5 &  2	\\

6593461	& 2001$\pm$18&  0.94$\pm$0.16 & 1.29$\pm$0.07   & 25.7$\pm$3.0 &  10.7$\pm$4.4 & 3 	\\

7296438$^\star$	& 1846$\pm$73&   1.10$\pm$0.02 & 1.37$\pm$0.01   & 25.2$\pm$2.8 &   6.4$\pm$0.6 & 2	\\

7680114$^\star$	& 1709$\pm$58&  1.09$\pm$0.03 &  1.40$\pm$0.01 & 26.3$\pm$1.9 &  6.9$\pm$0.5 &  2	\\

7700968	& 2010$\pm$25&  1.00$\pm$0.12 & 1.21$\pm$0.06   &  36.2$\pm$4.2 & 7.5$\pm$3.1 & 3	\\

9049593	& 1983$\pm$13&   1.13$\pm$0.14 & 1.40$\pm$0.06   & 12.4$\pm$2.5 &  6.4$\pm$3.4 &  3 	\\

9098294$^\star$	& 2347$\pm$84&  0.98$\pm$0.02 & 1.15$\pm$0.01  & 19.8$\pm$1.3 &  8.2$\pm$0.5 & 2 	\\

10130724	& 2555$\pm$27&  0.85$\pm$0.12 & 1.08$\pm$0.05   & 32.6$\pm$3.0 & 13.8$\pm$5.0 &   3 	\\

10215584	& 2172$\pm$28&  0.99$\pm$0.13 & 1.12$\pm$0.05   & 22.2$\pm$2.9 &  6.8$\pm$3.5 &  3 	\\

10644253$^\star$	& 2892$\pm$157& 1.09$\pm$0.09& 1.09$\pm$0.02  & 10.9$\pm$0.9 & 0.9$\pm$0.3& 2 	\\

10971974	& 2231$\pm$6&  1.04$\pm$0.12 & 1.09$\pm$0.03   & 26.9$\pm$4.0 & 5.8$\pm$3.0 & 3 	\\

11127479	& 1983$\pm$7&  1.14$\pm$0.12 & 1.36$\pm$0.06   & 17.6$\pm$1.8 &  5.1$\pm$2.2 &  3 	\\

11971746	& 1967$\pm$23&   1.11$\pm$0.14 & 1.35$\pm$0.06   &  19.5$\pm$2.1 &  6.0$\pm$2.8 &  3 	\\
\hline
\end{tabular}
\tablefoot{\Kepler input catalogue (KIC) number, central frequency of the oscillation power excess \num~\newer{by \cite{Chaplin2014}}, stellar mass and radius in solar units from seismology, surface rotation period P$_{\rm rot}$ from \cite{Garcia2014b},  stellar age from seismic modelling and reference to the published seismic studies:
[1]\,Garcia et al. \citep[in prep, see also][]{Beck2016}, 
[2]\,\cite{Creevey2016} and \hbox{[3]\,\cite{Chaplin2014}}.
Stars, for which the parameters were obtained through the detailed-modelling approach as described in \Section{sec:sample}, are flagged with an asterisk.
}
\label{tab:literatureParameters}
\end{table*}

In the last decade,  numerous studies focused on the question
whether the rotation and chemical abundances of the Sun are typical for a solar-type star,  
i.e. a star of a solar mass and age  \citep[e.g.][]{Gustafsson1998, AllendePrieto2006, DelgadoMena2014, Datson2014, Ramirez2014, Carlos2016, dosSantos2016}. These studies compared the Sun with solar-like stars and were inconclusive due to relatively large  systematic errors \citep{Gustafsson2008, Robles2008, Reddy2003}.

The fragile element lithium is a distinguished tracer of mixing processes and loss of angular momentum inside a star \citep{Talon1998}. Its abundance in stars changes considerably over the lifetime.  For low-mass stars during the main sequence, proton-capture reactions destroy most of the initial stellar Li content in the stellar interior. Only a small fraction of the original Li is preserved in the cool, outer convective envelopes.  
The solar photospheric Li abundance  A(Li) is 1.05$\pm$0.10\,dex \citep{Asplund2009}. This value is substantially lower than the protosolar nebular abundance of A(Li) = 3.3 dex \citep{Asplund2009}  measured from meteorites and illustrates that the lithium surface abundance does not reflect the star's original abundance.
A comparison between the Sun and typical stars of one solar mass and solar metallicity in the thin galactic disc by \cite{Lambert2004} showed the Sun  as ``lithium-poor'' by a factor of 10.
Because the temperature at the base of the solar convective zone is not hot enough to destroy lithium, this large depletion in the observed solar to the meteoritic Li abundance by a factor of 160 remains as one of the biggest challenges of standard solar models. This is known as the solar Li problem \cite[e.g.][and references therein]{Maeder2009,Melendez2010}.



There are two major challenges for understanding the Li abundance in stars similar to the Sun. First, for stars with masses M\,$\leq$\,1.1\,M\sun, there is a strong dependence of A(Li) on mass and metallicity, which are governed by the depth of the convective envelope for these stars \citep{doNascimento2009,doNascimento2010,Baumann2010,Castro2016}. While the stellar chemical composition can be determined from optical spectroscopy, the mass of a star is a difficult task when a star does not belong to a cluster. 
In this context, solar analogues, following the classical definition of \cite{CayreldeStrobel1996}, constitute a homogeneous set of stars in which mass and metallicity are well constrained with values close to solar ones as defined before. 
Fortunately, asteroseismology is a tool that provides precise and accurate 
values {for} mass, radius as well as ages {of} oscillating stars \citep*[e.g.][]{Aerts2010}. 
The highest level of accuracy of the parameters determined through seismology is reached when the models are constrained by individual frequencies and combined with results from high-resolution spectroscopy \citep[e.g.][]{Chaplin2011,Chaplin2014,LebretonGoupil2014}.

The second unknown is the complex interplay of various transport mechanisms and their efficiency inside the stellar interior and with stellar rotation. Standard models, that only include mixing through convective motion, fail to model the general trend of the A(Li) evolution. This indicates that additional mixing processes have to be taken into account, such as microscopic diffusion \citep{Eddington1916,Chapman1917}, inertial gravity waves \citep{GarciaLopez1991,Schatzman1996,Charbonnel2005} and the effects of stellar rotation. 
Rotation has a substantial impact on the stellar evolution \citep[e.g.][and references therein]{Zahn1992,MaederZahn1998,Brun1999,Mathis2004,Maeder2009,Ekstroem2012} and can change the properties of solar-type stars by reducing the effects of atomic diffusion, and inducing extra mixing. 
More specifically, observations of light element abundances bring precious constraints for mixing in models and transport processes in stars \citep{Talon1998,Charbonnel1998,Pinsonneault2010IAUS,Somers2016}.

\newer{Numerous observational and theoretical studies have explored the Li surface abundance in the context of rotation, stellar evolution, age and angular momentum transport \citep[e.g.][and references therein]{vandenHeuvel1971, Skumanich1972, Rebolo1988, Zahn1992, Zahn1994, Charbonnel1994, Talon1998, Charbonnel1998, King2000, Clarke2004,Talon2005, Pinsonneault2010IAUS, Bouvier2016, Somers2016}.} Recently, \cite{Bouvier2016} showed that a rotation-lithium relation exists already at an age of 5\,Myrs and also exhibits a significant dispersion. Moreover, \cite{Bouvier2008} proposed a possible link between lithium depletion and the rotational history of exoplanet-host stars.
Thus, authors seek a complete and coherent description of the influence of rotation on the lithium abundances on the main-sequence of solar-type stars.  
A particular challenge for 
studying A(Li) as a function of the stellar rotation 
is the surface rotation velocity. 
If determined from spectroscopy through the Doppler broadening of the absorption lines,  only the projected surface velocity \vsini could be measured, where the axis of the rotation axis remains unknown. If the surface rotation rate is determined through modulation of photometry or activity proxies, one measures the angular velocity in terms of surface rotation period, because a precise measure of the stellar radius is missing. 

\begin{figure}
\centering
\vspace{-3.25mm}
\includegraphics[width=0.9\columnwidth]{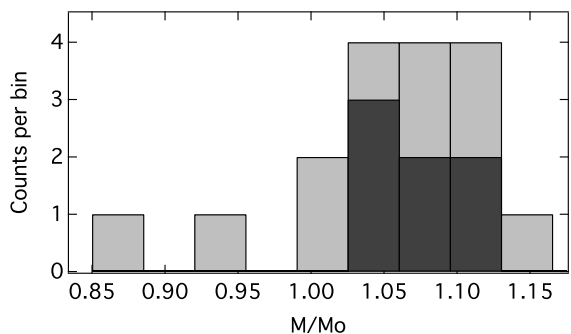}
\caption{Distributions of the stellar mass from seismology for the 18 solar-analogue stars in the presented sample. The grey bars flag the total distributions, while the dark shaded areas indicated the distribution of stars with detailed seismic modelling}
\label{fig:histograms}
\vspace{3mm}
\end{figure}

Understanding the evolution of the lithium abundance as a function of the mass, metallicity and rotation and explaining its dispersion in G~dwarfs,  is critical to construct a comprehensive model of the Sun as a star \citep[e.g.][]{Pace2012,Castro2016}.  Comparing the measured value of A(Li) in solar analogues with predictions from evolutionary models, calibrated onto the solar case will allow to test the evolution of the Li dilution for \blu{typical} 'Suns' at  \blu{different}  ages. This gives us the possibility to test if the mixing processes, assumed to act in the Sun\blu{,}  are peculiar or if the solar lithium value is  normal.

%


This paper is structured as follows: In \Section{sec:sample},  \blu{we}   set  \blu{the}  stars to be studied\blu{. Its} properties and the new spectroscopic parameters are described. From the observations, the relation between lithium and rotation is discussed in \Section{sec:lithiumRotation}. In \Section{sec:discussion}, the measured A(Li) is confronted with theoretical predictions \newer{from the Toulouse-Geneva stellar evolution code \citep[TGEC,][]{HuiBonHoa2008, doNascimento2009}} and \newer{we} compare age estimates from seismology \newer{derived from previous studies using different approaches}. Conclusions of this work are summarised in \Section{sec:Conclusions}.

\section{Data set and stellar parameters \label{sec:sample}}

\begin{figure}
\centering
\vspace{-1.75mm}
\includegraphics[width=1\columnwidth,height=120mm]{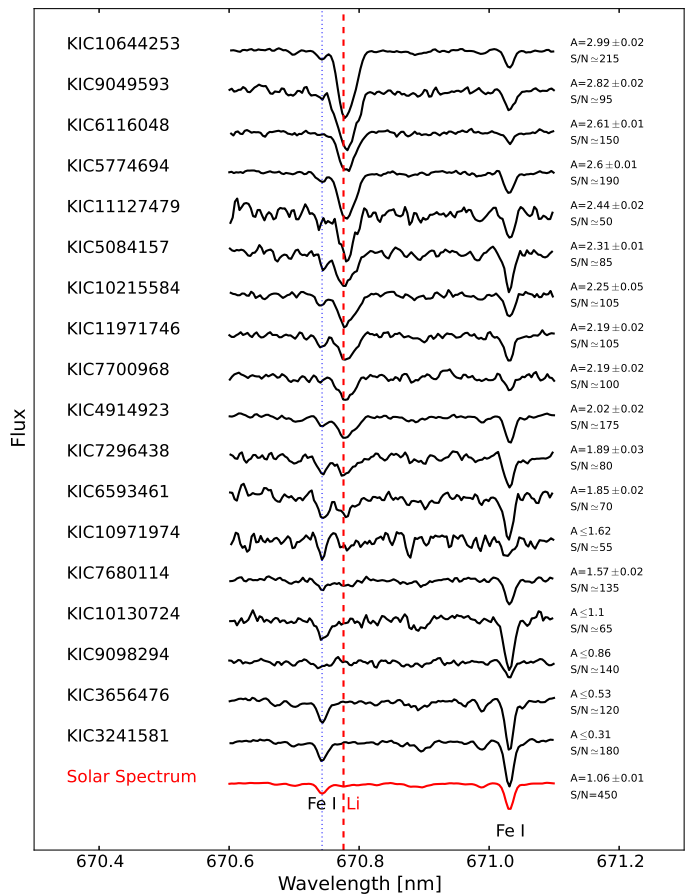}
\caption{Lithium doublet observed in the full dataset, sorted from strong to weak lithium lines (top to bottom, respectively). The solar spectrum (red) is shown in  the bottom of the diagram. The centre of the lithium \referee{and the neighbouring iron line are flagged} through the vertical dashed line \referee{and blue dotted line, respectively. The achieved S/N, measured lithium abundances as well as upper limits of A(Li) are indicated in the right side.}}
\label{fig:lithiumObservationalSequence}
\vspace{-3mm}
\end{figure}

The sample of solar analogues investigated in this study is composed by the 18 stars presented in \cite{Salabert2016Activity}. A summary of the main global properties of these stars are shown \hbox{in \Table{tab:literatureParameters}.}  
The stellar masses, radii and ages reported in the literature (\Table{tab:literatureParameters}) were obtained by either grid-modelling analysis of the global-seismic parameters or by using individual frequencies and high-resolution spectroscopy \newer{(hereafter also referred to as \textit{detailed} modelling)}. For a  \blu{specific}  discussion of the different modelling approaches\blu{,}  we refer the reader to \cite{LebretonGoupil2014}.  Detailed modelling using individual frequencies with the \textit{Asteroseismic Modeling Portal} \citep[\textsc{amp},][]{Metcalfe2009} is available for the following stars
KIC\,3656476, KIC\,4914923, KIC\,6116048, KIC\,7296438, KIC\,7680114, KIC\,9098294, and KIC\,10644253, which were modelled by \cite{Mathur2012}, \cite{Metcalfe2014} and \cite{Creevey2016}. \newer{In this paper we use whenever possible, the latest results of \cite{Creevey2016}.} An additional star, KIC\,3241581, has been modelled by Garcia~et~al. \citep[in prep., see also][]{Beck2016}, 
using the \textit{Modules for Experiments in Stellar Astrophysics} 
\citep[MESA,][and references therein]{MESA2013}.
For the remaining 10 stars, we adopted the masses and ages obtained by \cite{Chaplin2014} using global seismic parameters determined from 1-month long \Kepler time series and constraints on temperature and metallicity from multicolour photometry.  The mass distribution in \Figure{fig:histograms} shows that \blu{this} sample mainly consists of stars with masses in the upper half of the allowed mass regime for solar analogues (1-1.15\,M\sun). The dark shaded regions in \Figure{fig:histograms} depict the distribution of the stars for which detailed seismic modelling was performed \citep[Garcia et al., in prep.]{Mathur2012,Metcalfe2014,Creevey2016}. 
Whenever applicable, we distinguish in the diagrams  represented in this paper the values originating from the two analysis approaches. 

Surface rotation periods (P$_{\rm rot}$), \blu{measured by \cite{Garcia2014b}, are} reported in \Table{tab:literatureParameters}. {Selecting oscillating targets with known rotation periods adds \blu{a} constraint that these stars are magnetically active. However, they are not too active to suppress oscillations. This is another characteristic of our host star.}



\begin{table*}
\vspace{-1mm}
\caption{\blu{Summary of the seismic solar analogue observations.}}
\vspace{-1mm}
\centering
\tabcolsep=8pt
\begin{tabular}{rrrrrrrrrl}
\hline\hline
\multicolumn{1}{c}{KIC} & 
\multicolumn{1}{c}{V}&
\multicolumn{1}{c}{N} &
\multicolumn{1}{c}{ToT}&
\multicolumn{1}{c}{$\Delta$T}&
\multicolumn{1}{c}{$\overline{RV}$} & 
\multicolumn{1}{c}{$\Delta$RV} &
\multicolumn{1}{c}{comment}\\
& \multicolumn{1}{c}{[mag]} & &
\multicolumn{1}{c}{[hrs]} &
\multicolumn{1}{c}{[days]} &
\multicolumn{1}{c}{[km/s]} & 
\multicolumn{1}{c}{[km/s]} & 
  \\
\hline
3241581$^\star$	& 10.35$\pm$0.04	&24	&9.3	& 709.1	&-30.68 	&	0.96	&binary\\
3656476	& 9.55$\pm$0.02	&6	&2.5	& 351.1	&-13.23 	&	0.14			\\
4914923	& 9.50$\pm$0.02	&7	&2.1	& 297.3 	&-31.16 	&	2.11	&binary\\
5084157	& 11.56$\pm$0.12	&10	&5.2	& 300.1	&-19.66	&	0.21\\
5774694	& 8.37$\pm$0.01	&7	&1.3	& 348.1	&-17.67 	&	0.16	\\
6116048	& 8.47$\pm$0.01	&5	&1.0	& 347.2 	&-53.28	&	0.17	\\
6593461	& 11.22$\pm$0.10	&9	&4.1	& 296.2	&-35.39	&	[0.37]	& large scatter\\
7296438	& 10.13$\pm$0.03	&(+2) 4	&1.5	& 350.2	&-2.08	&	16.65	&Binary (KOI\,364.01)\\
7680114	& 10.15$\pm$0.04 	&8	&3.4	& 351.0	&-58.96	&	0.180	\\
7700968	& 10.37$\pm$0.04	&(+2) 4 	&1.4	& 299.1	&+39.47	&	27.62	&binary\\
9049593	& 10.35$\pm$0.04	&4	&1.5	& 299.1	&-21.02	&	0.24	\\
9098294	& 9.91$\pm$0.03	&7	&2.6	& 346.1	&-55.78	&	41.35	&binary\\
10130724	& 12.03$\pm$0.19	&7	&2.8	& 299.1	&-54.51	&	2.12	&binary\\
10215584	& 10.62$\pm$0.05	&6	&2.7	& 337.0	&-11.02	&	0.25	\\
10644253	& 9.26$\pm$0.02	&14	&5.9	& 416.0	&-19.01	&	0.18	\\
10971974	& 11.05$\pm$0.07	&4	&1.6	& 300.1	&-34.58	&	0.27\\
11127479	& 11.21$\pm$0.09	&5	&2.1	& 300.2	&-29.33	&	0.30	&KOI\,2792.01, large scatter	\\
11971746	& 11.00$\pm$0.07	&8	&3.6	& 301.0	&-44.20	&	0.22	\\
\hline
\end{tabular}
\tablefoot{The star's identifier in the \Kepler input catalogue (KIC)\blu{,}  apparent magnitude in Johnson\,V, number $N$ of spectra (in bracket the number of additional spectra with S/N only high enough to determine the radial velocity), total accumulated time on target (ToT),  the time base $\Delta$$T$ covered by the observations, the mean radial velocity ($\overline{RV}$), and the difference between the positive and negative extrema of the measured RV values and their internal error, and a comment on the star/system. KOI stand for \textit{Kepler Objects of Interests} and indicates planet host star candidates. }
\medskip
\label{tab:journalOfObservations}


\caption{Fundamental parameters of the solar analogues from the spectroscopic analysis  of \hermes data.}
\centering
\tabcolsep=10pt

\begin{tabular}{rlrrrrrr}
\hline\hline
\multicolumn{1}{c}{KIC} & 
\multicolumn{1}{c}{\teff} & 
\multicolumn{1}{c}{\logg} & 
\multicolumn{1}{c}{$v_{\rm min}$} & 
\multicolumn{1}{c}{\vsini} & 
\multicolumn{1}{c}{[Fe/H]} &
\multicolumn{1}{c}{A(Li)} &
\multicolumn{1}{c}{S/N} \\
\multicolumn{1}{c}{} & 
\multicolumn{1}{c}{[K]} & 
\multicolumn{1}{c}{[dex]} & 
\multicolumn{1}{c}{[km/s]} & 
\multicolumn{1}{c}{[km/s]} & 
\multicolumn{1}{c}{[dex]} &
\multicolumn{1}{c}{[dex]} &
\multicolumn{1}{c}{(Li)} \\
\hline

3241581	&	5685$\pm$59	&	4.3$\pm$0.1	&	1.0$\pm$0.2	&	4.0$\pm$0.6	&	0.22$\pm$0.04	&	\referee{$\leq$0.31}	&	180\\
3656476	&	5674$\pm$50	&	4.2$\pm$0.1	&	1.1$\pm$0.1	&	4.1$\pm$0.7	&	0.25$\pm$0.04	&	\referee{$\leq$0.51}	&	120\\
4914923	&	5869$\pm$74	&	4.2$\pm$0.1	&	1.2$\pm$0.1	&	5.0$\pm$0.6	&	0.12$\pm$0.04	&	2.02$\pm$0.02	&	175\\
5084157	&	5907$\pm$60	&	4.2$\pm$0.1	&	1.1$\pm$0.1	&	4.8$\pm$0.7	&	0.24$\pm$0.04	&	2.31$\pm$0.01	&	85\\
5774694	&	5962$\pm$59	&	4.6$\pm$0.1	&	1.0$\pm$0.2	&	5.5$\pm$0.7	&	0.10$\pm$0.03		&	2.60$\pm$0.01	&	190\\
6116048	&	6129$\pm$97	&	4.3$\pm$0.2	&	1.3$\pm$0.2	&	5.8$\pm$0.6	&	-0.18$\pm$0.05		&	2.61$\pm$0.01	&	150\\
6593461	&	5803$\pm$126	&	4.4$\pm$0.2	&	1.3$\pm$0.3	&	4.8$\pm$0.8	&	0.25$\pm$0.09		&	1.85$\pm$0.02	&	70\\
7296438	&	5854$\pm$64	&	4.3$\pm$0.1	&	1.2$\pm$0.2	&	4.5$\pm$0.8	&	0.24$\pm$0.05		&	1.89$\pm$0.03	&	80\\
7680114	&	5978$\pm$107	&	4.3$\pm$0.2	&	1.2$\pm$0.3	&	4.4$\pm$0.7	&	0.15$\pm$0.07		&	1.57$\pm$0.02	&	135\\
7700968	&	5992$\pm$144	&	4.4$\pm$0.3	&	1.4$\pm$0.4	&	5.3$\pm$0.7	&	-0.18$\pm$0.08		&	2.19$\pm$0.02	&	100\\
9049593	&	6009$\pm$151	&	4.3$\pm$0.3	&	1.5$\pm$0.3	&	7.2$\pm$0.6	&	0.20$\pm$0.08		&	2.82$\pm$0.02	&	95\\
9098294	&	5913$\pm$67	&	4.4$\pm$0.1	&	1.0$\pm$0.2	&	4.7$\pm$0.7	&	-0.14$\pm$0.04		&	\referee{$\leq$0.86}	&	140\\
10130724	&	5649$\pm$95	&	4.3$\pm$0.2	&	0.9$\pm$0.3	&	4.4$\pm$0.9	&	0.27$\pm$0.09		&	\referee{$\leq$1.10}	&	65\\
10215584	&	5888$\pm$67	&	4.3$\pm$0.1	&	1.1$\pm$0.2	&	5.1$\pm$0.6	&	0.05$\pm$0.04		&	2.25$\pm$0.05	&	105\\
10644253	&	6117$\pm$64	&	4.4$\pm$0.2	&	0.9$\pm$0.2	&	4.3$\pm$0.6	&	0.11$\pm$0.04		&	2.99$\pm$0.02	&	215\\
10971974	&	5895$\pm$114	&	4.4$\pm$0.2	&	1.2$\pm$0.3	&	4.8$\pm$0.8	&	0.02$\pm$0.07		&	\referee{$\leq$1.62}	&	55\\
11127479	&	5884$\pm$116	&	4.4$\pm$0.3	&	1.5$\pm$0.3	&	6.1$\pm$0.7	&	0.11$\pm$0.08		&	2.44$\pm$0.02	&	50\\
11971746	&	5953$\pm$63	&	4.3$\pm$0.1	&	1.2$\pm$0.2	&	4.5$\pm$0.8	&	0.18$\pm$0.04		&	2.19$\pm$0.02	&	105\\
\hline
\end{tabular}


\tablefoot{The star's identifier in the \Kepler input catalogue, the effective temperature \teff, the surface acceleration \logg, the micro turbulence $v_{\rm min}$, the projected surface rotational velocity \vsini, the stellar metallicity, and the abundance of lithium are given with their respective uncertainties. \referee{Upper limits of the measured A(Li) are indicated for stars with  low lithium abundance as consequence of insufficient S/N in the spectra.}  The last column reports the signal-to-noise ratio around the lithium line.}
\label{tab:fundamentalParameters}
\vspace{-2mm}
\end{table*}%

\subsection{Spectroscopic observations}
\begin{figure}
\centering
\vspace{-1mm}
\includegraphics[width=\columnwidth,height=50mm]{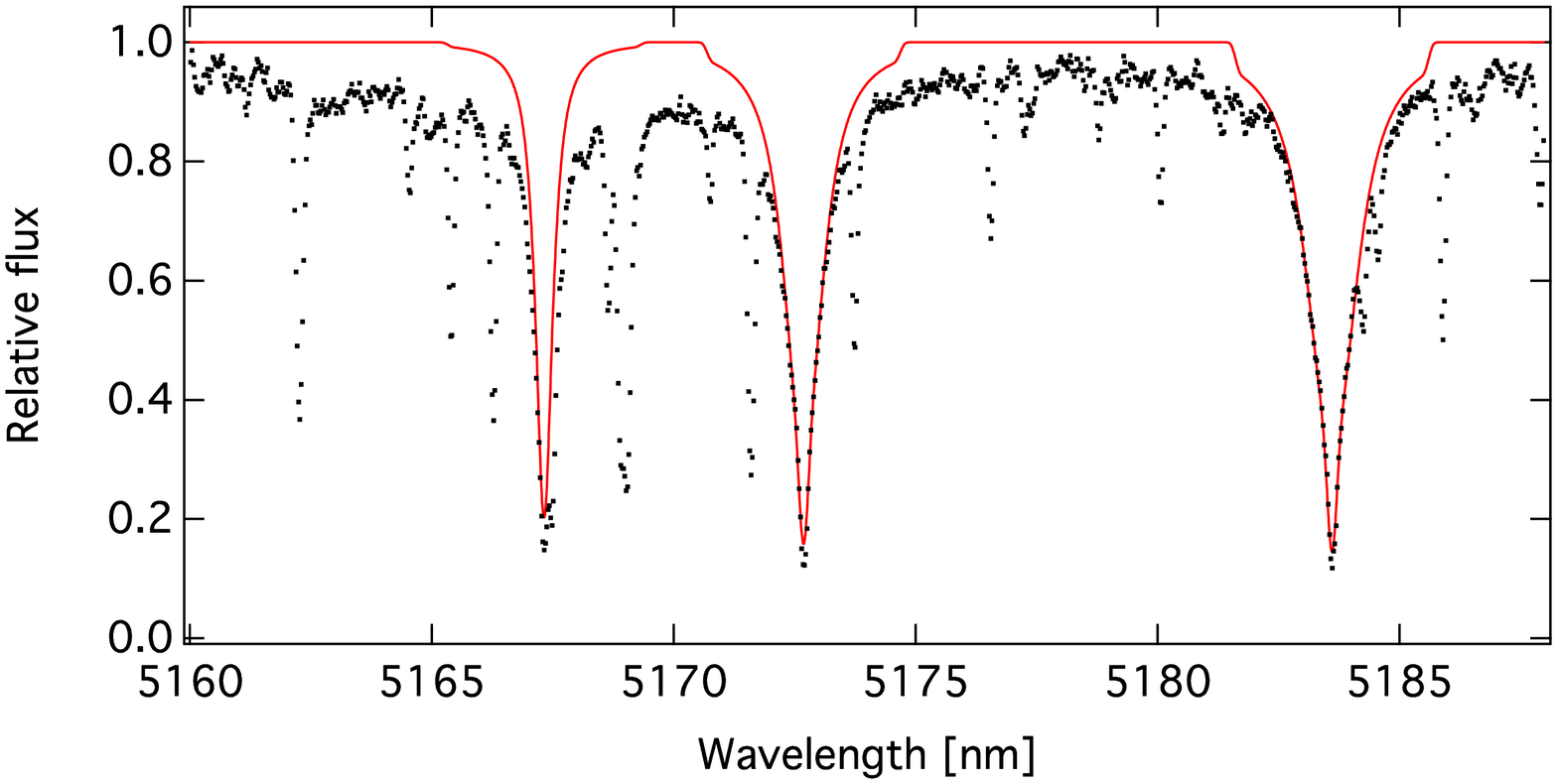}
\caption{Mg triplet in the star KIC\,7700968. The observed and synthetic spectra are represented by black dots and solid line, respectively. }
\label{fig:MgTriplet}
\end{figure}

To guarantee a homogenous sample of fundamental parameters, we obtained for each target high-resolution spectra with the \Hermes spectrograph \citep{Raskin2011,RaskinPhD}, mounted to the 1.2m \textsc{Mercator} telescope, La Palma, Canary Island, Spain. The observations were performed in four observing runs of 3 to 6 days each. In 2015, spectroscopic data was obtained in June and July, while in 2016 observations were obtained in April and May. The overview on the observations is presented in Table\,\ref{tab:journalOfObservations}.  In total 53.1\,hrs worth of exposure time were collected. The \Hermes  spectra cover a wavelength range between 375 and 900\,nm with a spectral resolution of R$\simeq$85\,000. The wavelength reference was obtained from emission spectra of Thorium-Argon-Neon reference frames in close proximity to the individual exposure.

The spectral reduction was performed with the instrument-specific pipeline \citep{Raskin2011,RaskinPhD}. The radial velocity (RV) for each individual spectrum was determined from the cross correlation of the stellar spectrum in the wavelength range between 478 and 653\nm with a standardised G2-mask provided by the \hermes pipeline toolbox.  For \Hermes, the 3$\sigma$ level of the night-to-night stability for the observing mode described above is $\sim$300\,m/s, 
which is used as the classical threshold for RV variations to detect binarity.  Using \cite{Beck2016} \blu{methods}  we corrected individual spectra for the Doppler shift before normalisation and the {process of combining} individual spectra. The {signal-to-noise ratios (S/N)} of each combined spectrum around 670\nm is reported in Table\,\ref{tab:fundamentalParameters} and   in  \Figure{fig:lithiumObservationalSequence}. A solar flux spectrum was observed with the same \hermes instrument setup in reflected sunlight from the Jovian moon Europa \citep{Beck2016}. This spectrum has a S/N\,$\sim$\,450 (\Figure{fig:lithiumObservationalSequence}).

\subsection{Fundamental parameters 
\label{sec:fundamentalParameters}}

To determine the fundamental parameters, we started with effective temperature \teff, surface gravity \logg, metallicity [Fe/H], and micro turbulence $v_{\rm min}$ and projected surface rotational velocity \vsini from \blu{an} analysis with the   \textit{Grid Search in Stellar Parameters} (\gssp\footnote{The GSSP package is available for download at https://fys.kuleuven.be/ster/meetings/binary-2015/gssp-software-package.}) software package \citep[][]{Lehmann2011,Tkachenko2012, Tkachenko2015}. The library of synthetic spectra were computed using the {\sc SynthV} radiative transfer code \citep{Tsymbal1996} based on the {\sc LLmodels} code \citep{Shulyak2004}.   Then,  we used the \Hermes  high-quality spectra    to determine  the final stellar  fundamental  parameters (\teff, \logg, [Fe/H]) and lithium abundance.  
We employed the  excitation/ionisation equilibrium balance technique to find the stellar parameters that produced consistent   abundances of Fe\,\textsc{i} and Fe\,\textsc{ii}, and by using the solar reference value as described by  \cite{Melendez2012},  and \cite{Ramirez2014}.    For all stars in the sample we determined the fundamental stellar parameters  by performing a differential excitation-ionization equilibrium from the abundances of Fe\,\textsc{i} and Fe\,\textsc{ii}, \blu{and by} using the solar value as a reference\blu{,} as described by \cite{Melendez2012}, \cite{Monroe2013}, \cite{Melendez2014a} and \cite{Ramirez2014}. 
We combined Kurucz atmospheric models  \citep{Castelli2004} with equivalent width (EW) measurements of Fe\,\textsc{i} and Fe\,\textsc{ii} and the 2014 version of the 1D LTE code MOOG \citep{Sneden1973}. The EW were determined from the automated code ARES \citep{Sousa2007}. We applied the same method for all stars in our sample, considering the same regions of continuum. Final spectroscopic parameters for the stars are given in \Table{tab:fundamentalParameters}. Formal uncertainties of the stellar parameters were computed as in \cite{Epstein2010} and \cite{Bensby2014}. 
The median metallicity for the sample is~0.15\,dex. We note that we find a higher temperature for KIC\,10644253, compared to the previous findings of \cite{Salabert2016Mowgli}. We adopt this new value, because the analysis was improved through increased \blu{observing} time and \blu{by applying} a differential analysis\blu{, with the} {\Hermes solar spectrum}. We adopt the values in \Table{tab:fundamentalParameters}. {For KIC\,3241581, we confirm the results reported previously by \cite{Beck2016}.}

\begin{figure}
\centering
\centering
\includegraphics[width=0.9\columnwidth]{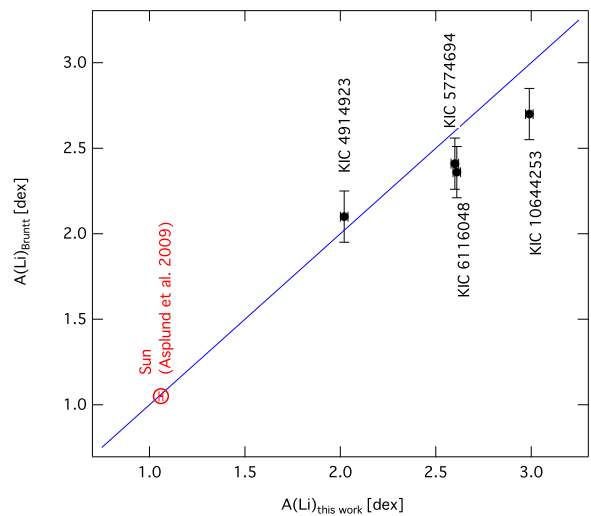}
\caption{Comparison of the measured lithium abundances with values \blu{from the literature} \referee{(see \Table{tab:LiLitComparison}). Black squares flag the four stars from our sample that are overlapping with the \cite{Bruntt2012}. The red circle depicts the comparison of the solar lithium abundance derived from our spectrum with the canonical value by \cite{Asplund2009}. The blue} line \blu{denotes} the 1:1 ratio between the two datasets.}
\label{fig:comparisonBrunt}
\vspace{3mm}
\end{figure}

The wings of Balmer and Mg-lines  in cool dwarfs stars are highly sensitive to the temperature, \logg and metallicity \citep{Gehren1981,Fuhrmann1993,Barklem2001}. These lines are formed in deep layers of the stellar atmosphere and they are expected to be insensitive to the non-LTE effects \citep{Barklem2007}, although they depend on convection \citep{Ludwig2009}.
The comparison between the observed and synthetic spectra for the  region between 516.0 and 518.8 \nm, containing the Mg-triplet and a sufficient number of metal lines, shown in \Figure{fig:MgTriplet}. The agreement of the \blu{line} widths shows the quality of our determined fundamental parameters (\teff, \logg, [Fe/H]).

\begin{figure*}
\centering
\includegraphics[width=\textwidth]{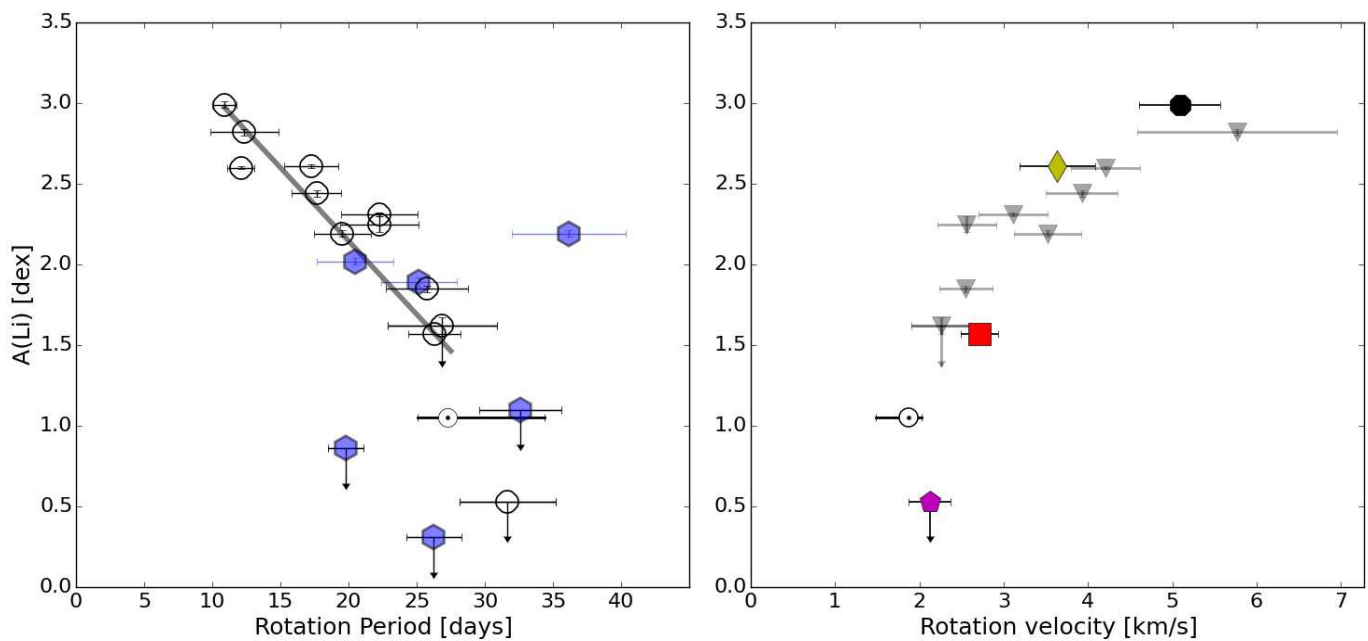}
\caption{{Lithium abundance versus rotation for the 18 seismic solar analogues. The left and the right panel compare A(Li) with the surface rotation period from space photometry and the computed surface rotation velocity, respectively. Stars found to be located in binaries are shown as filled blue \referee{octagons} in the left-hand panel and they are removed from the right panel.
 In both panels, the full range covered by the differential solar rotation is represented by the horizontal error bar in the solar symbol. 
The solid line in the left panel depicts the best fitting relation for single stars between the rotation period and A(Li) for stars with rotation periods shorter than the solar value.
In the right panel, the source of the asteroseismic radius is illustrated through the choice of symbols. Triangle markers indicate stars for which the radius has been determined through global-parameter modelling.  \referee{Upper limits of the measured A(Li) are shown for stars with  low lithium abundance or insufficient S/N in the spectra.} The single stars with the radius from detailed modelling from \textsc{amp},  KIC\,10644253, KIC\,6116048, KIC\,7680114, and KIC\,3656476 are plotted as black octagon, yellow diamond, red square and magenta pentagon, respectively.  }}
\label{fig:lithiumRotation}
\vspace{2mm}
\end{figure*}

\begin{table}
\caption{\referee{Comparison of lithium abundances with other values in the literature.}}
\centering
\begin{tabular}{rrrc}
\hline\hline
\multicolumn{1}{c}{Star} & \multicolumn{2}{c}{A(Li) [dex]}  & Literature \\
& This work & Literature & reference\\
\hline
KIC\,4914923	& 2.02$\pm$0.02 	& 2.1$\pm$0.2 	& B12\\
KIC\,5774694	& 2.6$\pm$0.01 	& 2.4$\pm$0.2 	& B12\\
KIC\,6116048	& 2.61$\pm$0.01 	& 2.4$\pm$0.2	& B12\\
KIC\,10644253	& 2.99$\pm$0.02 	& 2.7$\pm$0.2 	& B12\\\hline
Sun	& 1.06$\pm$0.01 	& 1.05$\pm$0.1 	& A09\\
\hline
\end{tabular}
\tablefoot{\referee{The stellar identifier, the lithium abundance derived in this work as well as in the literature value, provided by \cite{Bruntt2012} (B12) or \cite{Asplund2009} (A09). The comparison is depicted in \Figure{fig:comparisonBrunt}.}}
\label{tab:LiLitComparison}
\end{table}%

\subsection{Lithium abundance}
The A(Li) was derived from the Li I resonance doublet feature at 670.78\,nm as depicted for all stars in our sample in Figure\,\ref{fig:lithiumObservationalSequence}. We used the 'synth' driver of the 2014 version code MOOG \citep{Sneden1973} and adopted A(Li)$_\odot$ = 1.05 dex as the standard solar  lithium abundance  \citep{Asplund2009}. The atmosphere model  used were interpolated from the new Kurucz's grid \citep{Castelli2004} for \blu{a} set of spectroscopic atmospheric parameters, \teff, \logg, [Fe/H], and micro turbulence given in \Table{tab:fundamentalParameters}. We used the Fe\,\textsc{i} and Fe\,\textsc{ii} absorption lines \blu{as specified in} \cite{Melendez2014}, and we neglected possible $^6$Li influences.	
 Due to the vicinity of the Li lines to the Fe\,\textsc{i} line at 670.78\nm (\referee{blue dotted line in} Figure\,\ref{fig:lithiumObservationalSequence}), 
strong Li or iron lines as well as fast rotation can lead to blended lines.  Therefore, an accurate value of the iron abundance, the \logg, and the projected surface velocity is needed to correctly derive the lithium abundance. The main sources for the Li abundance error are related to the uncertainties on the stellar parameters and the EW measurement. However \teff is by far the dominant source of error.  For the spectroscopic atmospheric parameters\footnote{In this work, we use the standard definitions: $[X/Y]$=$\log$(N$_X$/N$_Y$)-$\log$(N$_X$/N$_Y$)$_\odot$, and $A_X$=$\log$(N$_X$/N$_H$)+12, where $N_X$ is the number density of element X in the stellar photosphere.},  we determined the lithium abundance in our sample ranging between 0.06 and 3.03\,dex. 
For comparison, the solar lithium abundance was \blu{also} derived from the \hermes solar spectrum (Figure\,\ref{fig:lithiumObservationalSequence}) collected  from the reflected light of the Jovian moon Europa. We measured A(Li)$_\odot$=1.06$\pm$0.1\,\dex in agreement with  \cite{Asplund2009}. The final values of A(Li) are listed in the last column of Table\,\ref{tab:fundamentalParameters} and Figure\,\ref{fig:lithiumObservationalSequence}. 
 \Figure{fig:lithiumObservationalSequence} also illustrates the sequence of spectral segments, containing the two lithium as well as two iron lines for all stars in our sample, sorted by  decreasing value of A(Li). For comparison, the solar spectrum obtained by \cite{Beck2016} was plotted at the bottom of the sequence. The comparison between our derived abundances with the values from \cite{Bruntt2012} is presented in \Figure{fig:comparisonBrunt}. A good agreement was found between the values for A(Li). The differences are probably originating from the NLTE effects in the hotter stars \citep{Lind2009}.

\subsection{Binarity occurrence}
The time span covered by our measurements, as well as the mean value and the dispersion of the radial velocities, are reported in Table\,\ref{tab:fundamentalParameters} for each~star in the sample. The measurements range between 260 and 700 days. Based on an earlier analysis of \KIC{3241581}, \cite{Beck2016} confirmed  this star to be a binary with \blu{an orbital} period longer than 1.5\,years. Based on the first 35 days of the observations of this campaign, 
 \cite{Salabert2016Activity} reported \KIC{4914923}, \KIC{7296438} and \KIC{9098294} as binary candidates. Additional spectra were needed to  confirm the binarity status of those systems. From the full available dataset analysed in this paper, we confirm the three above mentioned systems are binaries and we report that \KIC{10130724} and \KIC{7700968} are also binary systems. 

\blu{For none of the systems a binary period is known yet, because we do not detect the signature of stellar binarity (eclipses or tidally induced flux modulation) in their light curves. Also the RV measurements are too sparsely sampled to derive an orbital period from it.} 
Therefore no meaningful upper or even lower limit can be proposed on the orbital periods.  The mean value reported in Table\,\ref{tab:journalOfObservations} will roughly resemble  the systemic velocity of the binary system.   
\newer{Without information on the orbital parameters, the interpretation of A(Li) in the stellar components of the system is not reliable.} Therefore, continuous RV monitoring is required to draw further constraints on the orbital parameters, such as period or eccentricity.

\section{Lithium abundance and surface rotation \label{sec:lithiumRotation}}

There is a large number of observational works studying the connection between Li and rotation, searching for correlations between these parameters  \citep[e.g.][]{Skumanich1972,Rebolo1988,King2000,Clarke2004,Bouvier2008,Bouvier2016}.  {Because of the difficulty \newer{in coverage and stability} of photometric follow-up observations, most of them have employed \vsini measurements. Due to the undetermined inclination angle~$i$, such values yield a lower limit on the rotational velocity.} 
The rotation period from the light-curve modulation, \newer{such as determined by \cite{Krishnamurthi1998} for the Pleiades or from the modulation of the emission in the core of the Ca\,H\&K lines \citep[e.g.][]{Choi1995}}, is independent of the inclination. This, linked to the observational difficulty resolving the tiny absorption line of lithium at 670.7\nm  with high S/N for the solar-analogue stars, partially explains the difficulty to connect the dependence between true rotation (rotational period) and the lithium abundance in low-mass,  solar-analogues  stars at  different ages.


In the left panel of \Figure{fig:lithiumRotation}, the lithium abundance is plotted as a function of the surface rotation period derived from the \Kepler light curves \citep{Garcia2014b} and surface rotation velocity.  For comparison,  the Sun is presented by the longitudinal average rotation period of  27\,days.  The full range of solar differential rotation, spreads between  25\,days at the equator and  34 days at the poles \citep[e.g.][]{Thompson1996}, is represented by the horizontal error bar in the solar symbol in \Figure{fig:lithiumRotation}.  
From this figure we can see that  fast rotating stars have high Li abundances. 
This confirms the \blu{well known} general trend for lithium and rotation found in the \blu{earlier} mentioned studies of clusters and for single field stars \citep[e.g.][]{Skumanich1972,Rebolo1988,King2000,Clarke2004,Bouvier2008,Bouvier2016}.  
There is a large scattering \blu{for}  rotation periods longer than the solar rotation period. This is also found in similar studies of clusters and large sample of field stars.
In this context, our sample is unique in the sense that for field stars we combine the existing information about the \blu{true} rotation period, lithium abundance, seismic  \blu{age and} masses,  and binarity status.
In addition, stellar binarity can affect the measured lithium abundance either through interactions of the components \cite[e.g.][]{Zahn1994} or  due to observational biases. 
Given that an orbital solution is needed for a complete analysis of such system, we will concentrate on the single stars \blu{in this work}. This approach reduces the scatter in the A(Li)-\Prot plane, if only single stars are taken into account. 
In the rest of the analysis we only use single stars and stars with \prot below 27\,days.
For this subsample, the  Li-rotation correlation shows a trend following a linear regression,
\begin{eqnarray}
{\rm A(Li)}~=~-(0.08\pm0.01)\,\times\,{\rm P}_{\rm rot}~+~(3.85\pm0.17)\,.\label{eq:LiProt}
\end{eqnarray}

This relation indicates that lithium appears to evolve similarly to the rotation velocity for stars on this range of mass and metallicities \hbox{for \prot$\lesssim$\,27\,days}. 
We note that by fitting a trend in the \Prot-Li plane, we do not take explicitly into account the age of the stars. But implicitly, it is taken into account as surface rotation is a proxy of age \citep{Barnes2007}. Although \cite{vanSaders2016} showed that stars, once they reach approximately the age of the Sun, are not slowing down as much as predicted by the empirical relations \new{between rotation and stellar age} \citep[e.g.][]{Barnes2007,Gallet2013,Gallet2015}. \blu{However} gyrochronology is still valid for stars younger than the Sun. Thus, a rotation rate higher than the solar value implies a star younger than the Sun. This is in agreement with the modelled evolution of Li, because the strongest \blu{depletion} of the lithium surface abundance on the main sequence is taking place in the early stages \citep[e.g.][and references therein; see also \Section{sec:Conclusions} in this paper]{Castro2016}. 

The evolution of the lithium abundance with the rotation period, as described by \Equation{eq:LiProt} is the best fit of the trend for stars with masses 1.0\,$\lesssim$\,M/M$_\odot$\,$\lesssim$\,1.1 and [Fe/H] in the range 0.1 to 0.3\,dex, (with the exceptions of KIC\,10971974 and KIC\,6116048, \Table{tab:fundamentalParameters}).
The trend is well defined for rotation periods shorter than 27 days. The bulk of stars in our sample is rotating with shorter rotation periods than the Sun. This could be due to a selection bias in which longer periods are more difficult to detect. Besides that, more data seems to be necessary to extrapolate this results for slow rotators (long period) regimes.

Although, two stars \blu{with} the same angular velocity {(as used in the left panel of \Figure{fig:lithiumRotation}), the rotational velocity still can be different, since it depends on the unknown stellar radius.} We can overcome this degeneracy by using the asteroseismic radius (\Table{tab:literatureParameters}) to convert the rotation period into the unprojected rotational velocity $v_{\rm rot}$ in kilometers-per-second.
In the right panel of \Figure{fig:lithiumRotation},  we plot the rotational velocity computed  as a function of lithium abundance.  Because spots can be found at a relatively wide range of latitudes, surface differential rotation might contribute to the scatter in the rotation-lithium relation. 
\newer{Furthermore, the flux modulation introduced by spots, in combination with an assumed solar or anti-solar differential rotation profile will lead to an under or over estimation of the rotational velocity at the equator, respectively \newer{\citep[Brun et al., subm.]{Brun2015}}. From the right panel of \Figure{fig:lithiumRotation} we find that f}or stars with rotational velocities higher than 3 km\newer/s a linear trend is found in the Li-velocity relation. 
Between values of $v_{\rm rot}$ of 2 and 3 km/s, a large dispersion in the measured values of lithium could be present (A(Li)$\lesssim$2\,dex),  engulfing also the position of the Sun in this diagram. \newer{Comparing the right panel of \Figure{fig:lithiumRotation} to large-sample studies of \vsini, such as \cite{Takeda2010} shows a good agreement. In the asteroseismic approach, the complications introduced by the unknown inclination of the rotation axis or the assumptions on the turbulent velocity fields that could influence the line profile, are eliminated. Applying this asteroseismic approach on large samples should thus reduce the systematic scatter.}

\section{Discussion \label{sec:discussion}}

Both observables, rotation and A(Li) are expected {to evolve with time} for stars of this mass range during the main sequence. This was suggested by \cite{Skumanich1972} from the observations of stars in the Hyades, Pleiades, Ursa Major and the Sun, which were further investigated \citep[e.g.][]{Rebolo1988, King2000,Clarke2004}. 
%
 In general, the Li abundance is a function of the  convective envelope deepening relative to the age of a star on the main sequence \citep{doNascimento2009,Baumann2010}. {It can also be the consequence} of mixing below the convective zone \citep{Brun1999} \new{or in the radiative core \citep{Talon2005,Charbonnel2005}}.  
 {This confirms that the age, the angular momentum history, the mass and the metallicity are the} governing physical processes in the evolution of the lithium content. Recently,  \cite{Castro2016} showed that in the cluster \object{M67}, there is a relatively large scatter of the lithium abundance in the main-sequence stars with the same effective temperature and same age. The scatter is the largest around the 1\,M\sun-range (0.5\,$\lesssim$\,A(Li)\,$\lesssim$\,2.0\,dex) and suggests that another, yet unknown process could have an influence on the lithium abundance. \cite{Somers2016} suggest that an intrinsically different mixing history \new{than in other stellar clusters, such as a higher fraction of fast rotators or inhomogeneities of the initial rotation distribution \citep{Coker2016}}, could explain the scatter in lithium depletion \hbox{for stars older than 100\,Myr.}

\subsection{Stellar ages}
In a cluster all stars have the same age. This cannot be assumed for \blu{the field} stars in our sample.
{As described in \hbox{\Section{sec:sample}}, there are two ways to use the seismic information to infer the age. When grid-modelling based on global-seismic parameters is used, the inferred ages have large uncertainties \citep{LebretonGoupil2014}. 
In our sample, several of these stars are rotating faster than the Sun \citep{Garcia2014b}, \hbox{indicating} that they should be younger {\citep{Barnes2007, vanSaders2016}}. However, \blu{some} ages \blu{are} derived from global-parameter seismology\blu{, and} are larger than the solar age.
To avoid these \blu{inconsistency} problems, in the following analysis, we will only use ages inferred from detailed modelling, constrained by individual frequencies or frequency ratios.} 
\new{As described in \Section{sec:sample} and listed in \Table{tab:literatureParameters}. We have age and mass estimates through this approach for eight stars from studies of \cite{Mathur2012}, \cite{Metcalfe2014}, {\cite{Creevey2016} and Garcia et\,al.\ (in\,prep.)}.}

\subsection{Comparison with stellar models}

To compare the derived stellar age and measured lithium abundance with stellar modelling predictions, a grid of models of the temporal evolution of A(Li) due to rotation-induced mixing using the \textit{Toulouse-Geneva stellar evolution code} \citep[TGEC,][]{HuiBonHoa2008, doNascimento2009} was computed. 

A description of \referee{the} physics used for this grid is given in Appendix\,\ref{sec:TGEC}. {For details on the calculation of the theoretical Li abundance we refer to \cite{Castro2009,Castro2016}, and \cite{doNascimento2009}. } \referee{These models include {the impact of the} rotation-induced mixing {on chemicals due to the combined actions of meridional circulation and shear-induced turbulence in stellar radiation zones} 
{computed} as prescribed by \cite{Zahn1992}, \cite{MaederZahn1998}, \cite{TheadoVauclair2003} and \cite{HuiBonHoa2008}. This non-standard mixing, modelled as a vertical {effective turbulent} diffusion {applied on} chemical elements \citep[][and Appendix\,\ref{sec:TGEC}]{Chaboyer1992}, is calibrated to reproduce the solar lithium abundance at the solar age in a solar model \citep[we refer the reader to][for a detailed description of the calibration]{Castro2016}.} The calibration is then used for the other models with different masses and metallicities. 
{In this framework, it is important to point that we focus here only on the non-standard mixing for chemicals (we refer the reader to Appendix\,\ref{sec:TGEC} for more details). This rotational mixing and the resulting Lithium abundance strongly depend on internal transport mechanisms \citep[e.g.][]{Maeder2009,Mathis2013LNP} and on angular momentum extraction by stellar winds \citep[][]{Skumanich1972,Mattetal2015} as illustrated for exemple by the recent work by \cite{Amardetal2016}. These processes are here calibrated as explained in Appendix\,\ref{sec:TGEC} on the Sun and its rotational history through the effective vertical turbulent diffusion acting on chemicals. This may introduce a bias towards solar characteristics. 
However this work constitutes a first step. In a near future, more realistic models will be computed where rotational and chemical evolutions will be treated coherently and simultaneously. These models will take into account all angular momentum processes and potentially different rotational histories.}

\begin{figure}
\centering
\includegraphics[width=1\columnwidth,height=60mm]{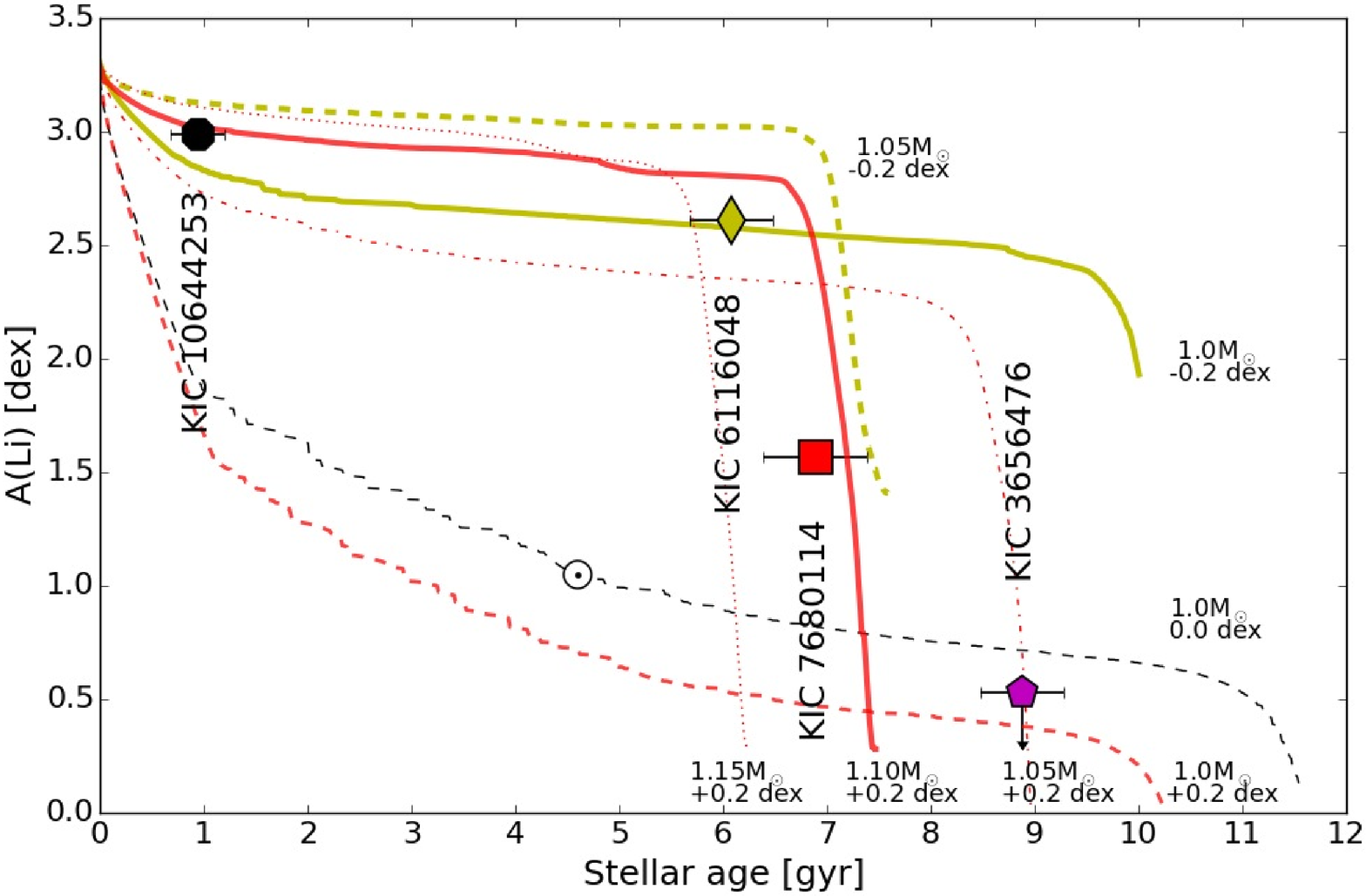}
\caption{Lithium abundance \blu{of} single stars with \blu{age computed by} detailed modelling \new{ with the \textsc{amp} code. 
 The stars KIC\,10644253, KIC\,6116048, KIC\,7680114, and KIC\,3656476 are plotted as black hexagon, yellow diamond, red square and magenta pentagon, respectively.} \referee{For KIC\,3656476 the upper limits of the measured A(Li) is shown.}  {TGEC-}evolutionary tracks shown in red, and yellow represents the theoretical evolution of A(Li) for models of the indicated mass with [Fe/H]=+0.2, and $-0.2$\,dex, respectively. 
The black dashed evolutionary track depicts the evolution of Li calculated by \cite{doNascimento2009} 
and the solar marker depicts measured A(Li\sun) and age of the Sun.}
\label{fig:LithiumAge}
\end{figure}

The setup of the input physics used in {our} models is compatible with the one used in the \textsc{amp} models. Our models are calculated from the zero-age main sequence (ZAMS) to the helium burning in the core and for masses from 1.0 to 1.15 M$_\odot$ with a step size of 0.05\,M\sun.   {The main grid was calculated with a metallicity of [Fe/H]=$+$0.20.}
\new{An additional model of} {the Li-evolution was calculated for a 1.0\,M\sun star with [Fe/H]=$-$0.20.
For comparison, the evolutionary track of the solar model computed by \cite{doNascimento2009} is shown in \Figure{fig:LithiumAge}.}

We note that the grid of models contains only representative models chosen for average mass and metallicity. The step size of 0.05\,M$_\odot$ is roughly the averaged uncertainty (7\%) of the detailed modelling approach found by \cite{LebretonGoupil2014}, accounting for the observational uncertainties and those of the model approaches. The authors found also a typical uncertainty of the seismic age estimate of $\sim$10\%.
Therefore, this comparison provides a qualitative idea if the models and the measurements agree in general, but these model tracks are not calibrated to resemble specific stars on a case-to-case basis.


\new{By focusing on stars for which detailed modelling has been performed {and were found to be without a companion}, our sample is narrowed down to 4 stars, \blu{they are} $-$} 
\new{
KIC\,3656476 (age=8.9\,Gyr, [Fe/H]=+0.25\,dex), 
KIC\,6116048 (6.1\,Gyr, $-0.18$\,dex),  
KIC\,7680114 (6.9\,Gyr, +0.15\,dex),  and 
KIC\,10644253 (0.9\,Gyr, +0.1\,dex). 
 The measured lithium abundances 
 are compared against the stellar age from \textsc{amp}-modelling \blu{for} these 4 stars in \Figure{fig:LithiumAge}.
\blu{Here, these} three stars KIC\,10644253, KIC\,7680114 and KIC\,3656476, are of particular interest, as they form a sequence of constant mass of 1.1\,M\sun (within their uncertainties) and  a metallicity above \blu{the} solar \blu{value}, which spans over stellar ages between 1 and 9\,Gyr. With its 1.05$\pm$0.03\,M\sun, KIC\,6116048 is closer to the solar mass, but has a clear sub-solar metallicity. In \Figure{fig:LithiumAge} 
the observed A(Li) 
is plotted as a function of the estimated age from asteroseismology and compared to the predictions of these quantities from the above described model tracks.}

\subsubsection{Case of KIC\,10644253}

\new{The comparison depicted in \Figure{fig:LithiumAge} 
shows that the best agreement between measurements and models for age and lithium 
is found for KIC\,10644253.  The measured activity levels as well as the short rotation period reported by \cite{Salabert2016Activity,Salabert2016Mowgli} further confirm that this \blu{is a young} stellar object.}\vspace{-1mm}

\subsubsection{Case of KIC\,7680114}
{KIC\,7680114 complies reasonably well with the evolutionary track of a star of 1.1\,M\sun with a metallicity of +0.2\,dex. {The star has a rotation period of 26.3\,days, compatible with the solar one. The} 
asteroseismic modelling places it at $\sim$7\,Gyr, {about 2.5\,Gyr older than the Sun.} 
\cite{vanSaders2016} show that when increasing mass or temperature the reduced efficiency of magnetic breaking starts earlier, which may explain this discrepancy.}\vspace{-1mm}

\subsubsection{Case of KIC\,3656476}
The slowly rotating star KIC\,3656476, is confirmed from stellar modelling and the comparison with the lithium abundance to be an old object. 
Although the 1.1\,M\sun evolutionary track of the TGEC modelling does not reach the age predicted by the  \textsc{amp} model it is not a worrying disagreement. 
Taking the typical uncertainty of $\sim$7\% on the seismic mass estimate, we find a good agreement with the Li evolution for a 1.05\,M\sun star.
Such uncertainty corresponds to 2\,$\sigma$ of the mass uncertainty of the star's model reported by \cite{Creevey2016}. \vspace{-1mm}



\subsubsection{Case of KIC\,6116048}
For KIC\,6116048, the rotation period of $\sim$10 days shorter than the value of the Sun suggests that this is a young star \citep{Garcia2014b}. On the other hand, this star is one of the \blu{lowest} active stars in our sample \citep{Salabert2016Activity} and the seismic modelling suggests an age of 6\,Gyr \citep{Creevey2016}.
In principle, it is possible that the rotation period from the literature could be half of the actual value \citep{Garcia2014b}. In such case however, this star would fall out of the observed relation between \Prot and A(Li.)

Despite the non-agreement of the age indicators, we find a general good agreement \blu{between} A(Li) measurement and the models. Within the conservative view on the uncertainty of the mass, the measured A(Li) is in good agreement with the \blu{theoretical} Li evolution for a 1.0 and 1.05\,M\sun star. \new{Because A(Li)  remains relatively constant over a large range in time, Li cannot be used to distinguish between the two scenarios. KIC\,6116048 is a puzzling case which definitely needs further investigation.} \vspace{-1mm}
 
 \subsection{Results}
{From the comparison of the observations with models, calculated for the determined seismic mass, we find A(Li) in good agreement for all \blu{four} single stars with available \textsc{amp} modelling. }
For the Sun it has been shown that gravity waves have to be included in order to reproduce the solar rotation and lithium profile \citep{Charbonnel2005}. {Although gravity waves were not explicitly included in the applied macroscopic \referee{hydrodynamical} 
mixing modelling, they were implicitly taken into account through the calibration of the models to the Sun. }
{The good agreement of measured and modelled A(Li) in \Figure{fig:LithiumAge} shows that the four stars share the same internal physics as it is working in the Sun.}



\section{Conclusions \label{sec:Conclusions}}

In this work, we have presented the combined analysis of seismic, photometric, and spectroscopic values for a set of 18 solar-analogue stars. 
The sample is \blu{important and} unique, as not only \blu{for} the lithium abundance and the rotation period are known, but also \blu{for the} mass, radius and age estimates for all stars \blu{from seismology.}
The rotation periods and seismic parameters used in this study were determined by several earlier studies \newer{\citep[][Garcia et al. in prep.]{Mathur2012,Garcia2014b,Chaplin2014,Metcalfe2014,Creevey2016}}.  \newer{For an overview of the literature values and references please see \Table{tab:literatureParameters}.} In a dedicated observing campaign we obtained high-resolution spectroscopy with the \hermes \'echelle spectrograph, allowing us to determine a consistent set of spectroscopic fundamental parameters, including the \blu{Li} surface abundance.  From our spectroscopic observations, \blu{we detected} six new binary systems.

The surface abundance of lithium of a star is very sensitive to {its rotation rate, mass, and metallicity.} \blu{M}asses from asteroseismology allows us to select our targets accurately on the criterion of mass.  Choosing a sample of stars within the very narrow mass range, which is accepted \blu{as a} solar-analogues, enables us to study the interplay of these parameters.  
 From the sample of single stars, we could quantify a linear relationship between the rotation period and the A(Li) for rotation periods shorter than the solar one. Binary stars show a larger scatter in the parameter plane.
We demonstrated that  observational restriction can be overcome by calculating the actual rotational velocity $v_{\rm rot}$ using the asteroseismically determined radius of the star. This allows a  better comparison with model predictions.

\newer{By focusing on 4 single stars with available masses and \newer{robust} ages from detailed modelling with the \textsc{amp}-code and confronting them with TGEC evolutionary track for the lithium abundance, we confirm the high degree of  sensitivity of A(Li) to the combination of stellar mass, metallicity, rotation rate and age.} \newer{For \newer{two of the} 'massive' solar analogues ($\sim$1.1\,M\sun)  with detailed modelling, KIC\,10644253 and KIC\,7680114, the measured A(Li) and the stellar mass and age from asteroseismology agree well}  with the predicted Li abundance.  For the third 'massive' solar analogue, KIC\,3656476, a good agreement is found within the conservative mass uncertainty of $\sim$7\%. 
A similar case is the solar-analogue with sub-solar metallicity, KIC\,6116048. Also for this star we find a good agreement with the modelled evolution of A(Li) within the conservative mass uncertainty.
The measured value of A(Li) agrees with the plateau value found for 1.0\,M\sun star but the rotation period indicates a young object while seismology suggest a target, older than the Sun. In principle the rotation period could be underestimated by a factor of two, which however would lead to a strong outlier in the \prot-A(Li) relation, depicted in \Figure{fig:lithiumRotation}. 

The comparison of A(Li) with age and the rotation rate demonstrates that gyrochronology is valid for stars younger than the Sun
 \blu{and until the age of the Sun}. The small number of stars with individual-frequency modelling does not allow us to draw further conclusions on their evolution with age. A larger dataset will be required to confirm the conclusions outlined here.

\blu{For these genuine solar analogues, a} good agreement within the uncertainties is found between three independent approaches $-$ the observed A(Li) from spectroscopy, the stellar age and mass from asteroseismology as well as the stellar model prediction of A(Li) for representative TGEC-models. Because the TGEC-models for A(Li) were calibrated to reproduce the solar internal mixing, such consensus with the measured A(Li) in the solar analogues {may} indicate that the solar analogues share the same acting internal mixing than the Sun. In this light, the solar value of A(Li) \blu{is absolutely} normal.  \vspace{7mm}


\begin{acknowledgements}
\blu{We thank the referee for his/her constructive report that allows us to improve the article.} 
We acknowledge the work of the team behind \Kepler and \textsc{Mercator}.  
PGB and RAG acknowledge the ANR (Agence Nationale de la Recherche, France) program IDEE (n$^\circ$ ANR-12-BS05-0008) "Interaction Des Etoiles et des Exoplanetes". PGB and RAG also received funding from the CNES grants at CEA. JDN, MC and TD acknowledge the CNPq and the PPGF/UFRN. PGB acknowledges the PPGF/UFRN for founding partially a scientific visit to the G3 team at UFRN, Natal, Brazil. DS and RAG acknowledge the financial support from the CNES GOLF and PLATO grants. DM acknowledges financial support from the Spanish Ministerio de Econom{\'i}a y Competitividad under grant AYA2014-54348- C3-3-R.  StM acknowledges support by the ERC through ERC SPIRE grant No. 647383.
AT received funding from the European Research Council (ERC) under the European Union's Horizon 2020 research and innovation programme (grant agreement N$^{\circ}$670519: MAMSIE). SaM would like to acknowledge support from NASA grants NNX12AE17G and NNX15AF13G and NSF grant AST-1411685.
The research leading to these results has received funding from the European Community's Seventh Framework Programme ([FP7/2007-2013]) under grant agreement No. 312844 (SPACEINN) and under grant agreement No. 269194 (IRSES/ASK). The observations are based on spectroscopy made with the \Mercator Telescope, operated on the island of La Palma by the Flemish Community, at the Spanish Observatorio del Roque de los Muchachos of the Instituto de Astrof{\'i}sica de Canarias. 
This research has made use of the SIMBAD database, operated at CDS, Strasbourg, France.
\end{acknowledgements}

\bibliographystyle{aa}
\bibliography{bibliographyLithium.bib}

\begin{appendix}
\section{Physics of the TGEC-grid \label{sec:TGEC}}
\blu{A} grid of models was calculated \blu{with} the \textit{Toulouse-Geneva stellar evolution code} \citep[TGEC,][]{HuiBonHoa2008, doNascimento2009}. 
The initial chemical mixture relative to the hydrogen content was chosen as the solar one from \cite{Grevesse1993}.
The equations of state are derived from the OPAL tables \citep{Rogers2002} and we used the OPAL96 opacities tables \citep{Iglesias1996} completed by the \cite{Alexander1994} low temperature opacities. For the nuclear reaction rates, we used the NACRE compilation \citep{Angulo1999} with the Bahcall screening routine. 

Convection \blu{was} modelled according to the \cite{BoehmVitense1958} formalism of the mixing-length theory. The mixing-length parameter $\alpha = l/H_p = 1.69$, where $l$ is the characteristic mixing length and $H_p$ is the pressure scale height, is a free parameter which calibrates the radius of a solar model. Below the convective zone, we introduce a convective undershooting with a depth of 0.09 $H_p$ so that, in a solar model, the combined mixing reaches the depth deduced by helioseismology  \citep[$r_{cz}/R_{\odot}=0.713 \pm 0.001$;][]{BasuAntia1997}. In the radiative zone, microscopic diffusion, which is the process of element segregation by gravitational and thermal diffusion \citep{Eddington1916, Chapman1917}, is treated by using the \cite{Paquette1986} method for collisions between charged ions with a screened coulomb potential. For a complete description of the microphysics used by these models, we refer the reader to \cite{TheadoVauclair2003,doNascimento2009,Castro2016}.

The meridional circulation {in stellar radiation zones} \citep[e.g.][]{Eddington1926,Sweet1950}, which is driven by the internal stresses induced by rotation \citep[e.g.][]{Zahn1992,Rieutord2006,Mathis2013LNP}, is modelled as prescribed by \citet{Zahn1992}. \cite{Zahn1992} demonstrated that {these} meridional flows transport angular momentum, creating shears that become unstable {with a stronger turbulent transport in the horizontal direction than in the vertical one because of the stable stratification}.
{As demonstrated by \cite{Chaboyer1992}, the vertical advection of chemicals by the meridional circulation is transformed into an effective vertical diffusion because of the strong horizontal turbulence}.
{Similarly, } the transition layer at the bottom of the convection zone between its latitudinal differential rotation and the solid-body rotation of the radiative core (the so-called tachocline) undergoes the same strong anisotropic turbulence (with a much stronger turbulent transport in the horizontal than in the vertical direction). The resulting turbulent transport reduces the differential rotation and inhibits its spread deep inside the radiative interior \citep{SpiegelZahn1992}. \citet{Brun1998}  showed that the vertical turbulent transport of chemicals in these layers can be modeled by an exponential diffusion coefficient, which is added to the previously described effective vertical turbulent diffusion. The resulting total vertical turbulent diffusion coefficient in the transport equation for the mean concentration of the different chemical species is here calibrated to reproduce the solar lithium abundance in a solar model as in \cite{Castro2016}. In our models, we solve only the transport equation for chemicals. The angular momentum history is not computed explicitly since the equation for the transport of angular momentum is not solved in our models, but it is implicitly taken into account through calibration constants of the effective vertical diffusion, which are implicit function of rotation and shear \citep{Chaboyer1992,Zahn1992,HuiBonHoa2008,Castro2016}.


\end{appendix}

\end{document}